\documentclass{article}
\usepackage{emulateapj,graphicx,epsf,natbib}
\def\myputfigure#1#2#3#4#5%
{\hskip0.03\textwidth
\vskip#5pt
\makebox[0pt]{\hskip#2in
\includegraphics[width=#3\textwidth]{#1}}\vskip#4pt\hfill}

\newcommand{\uv}{\mbox{$u$-$v$}}
\newcommand{\Ho}{\mbox{$H_\circ$}}
\newcommand{\etal}{\mbox{et~al.}}
\newcommand{\no}{\mbox{$n_{e  \circ}$}}

\newcommand{\dTo}{\mbox{$\Delta T_\circ$}}
\newcommand{\Xo}{\mbox{$S_{x  \circ}$}}
\newcommand{\Teo}{\mbox{$T_{e  \circ}$}}
\newcommand{\Lamo}{\mbox{$\Lambda_{e \mbox{\tiny H} \circ}$}}
\newcommand{\LameH}{\mbox{$\Lambda_{e \mbox{\tiny H}}$}}
\newcommand{\Lamee}{\mbox{$\Lambda_{e e}$}}
\newcommand{\fx}{\mbox{$(x \frac{e^x+1}{e^x-1} -4)
	(1+\delta_{\mbox{\tiny SZE}})$}}
\newcommand{\Om}{\mbox{$\Omega_M$}}
\newcommand{\Ol}{\mbox{$\Omega_\Lambda$}}

\newcommand{\kms}{\mbox{km s$^{-1}$}}
\newcommand{\ksM}{\mbox{km s$^{-1}$ Mpc$^{-1}$}}
\newcommand{\cgsunits}{\mbox{$\frac{\mbox{erg s}^{-1} 
	\mbox{ cm}^{-2}}{\mbox{cnt s}^{-1}}$}}
\newcommand{\ecs}{\mbox{erg cm$^{3}$ s$^{-1}$}}

\newcommand{\kB}{\mbox{$k_{\mbox{\tiny B}}$}}
\newcommand{\sigT}{\mbox{$\sigma_{\mbox{\tiny T}}$}}
\newcommand{\Tcmb}{\mbox{$T_{\mbox{\tiny CMB}}$}}
\newcommand{\muH}{\mbox{$\mu_{\mbox{\tiny H}}$}}
\newcommand{\nH}{\mbox{$n_{\mbox{\tiny H}}$}}
\newcommand{\NH}{\mbox{$N_{\mbox{\tiny H}}$}}
\newcommand{\Da}{\mbox{$D_{\!\mbox{\tiny A}}$}}
\newcommand{\h}{\mbox{$^{\mbox{h}}$}}
\newcommand{\m}{\mbox{$^{\mbox{m}}$}}
\newcommand{\s}{\mbox{$^{\mbox{s}}$}}

\newcommand{\gsim}{\gtrsim}
\newcommand{\lsim}{\lesssim}

\newcommand{\clusterfull}{MS $0451.6-0305$}
\newcommand{\clustwofull}{Cl $0016+16$}
\newcommand{\cluster}{MS 0451}
\newcommand{\clustwo}{Cl 0016}

\begin{document}

\title{Sunyaev-Zel'dovich Effect Derived Distances to the
High-Redshift Clusters \clusterfull\ and \clustwofull}


\author{Erik~D.~Reese\altaffilmark{1},
Joseph~J.~Mohr\altaffilmark{1,2}, John~E.~Carlstrom\altaffilmark{1},
Marshall~Joy\altaffilmark{3}, Laura~Grego\altaffilmark{4},
Gilbert~P.~Holder\altaffilmark{1},
William~L.~Holzapfel\altaffilmark{5},
John~P.~Hughes\altaffilmark{6,7}, Sandeep~K.~Patel\altaffilmark{3,8},
Megan~Donahue\altaffilmark{9}}


\altaffiltext{1}{Department of Astronomy \& Astrophysics, University
of Chicago, 5640 S.\ Ellis Ave., Chicago, IL 60637}

\altaffiltext{2}{Chandra Fellow}

\altaffiltext{3}{Space Science Laboratory, SD50, NASA Marshall Space
Flight Center, Huntsville, AL 35812}

\altaffiltext{4}{Harvard-Smithsonian Center for Astrophysics, 60
Garden St., Cambridge, MA 02138}

\altaffiltext{5}{Department of Physics, University of California,
Berkeley, CA 94720}

\altaffiltext{6}{Department of Physics and Astronomy, Rutgers
University, 136 Frelinghuysen Road, Piscataway, NJ 08854-8019}

\altaffiltext{7}{Service d'Astrophysique, L'Orme des Merisiers,
Batiment 709, Commissariat \`{a} l'Energie Atomique-Saclay, 91191
Gif-sur-Yvette Cedex, France}

\altaffiltext{8}{Department of Physics, University of Alabama,
Huntsville, AL 35899}

\altaffiltext{9}{Space Telescope Science Institute, 3700 San Martin
Dr., Baltimore, MD 21218}

\begin{abstract}
We determine the distances to the $z \simeq 0.55$ galaxy clusters
\clusterfull\ and \clustwofull\ from a maximum likelihood joint fit to
interferometric Sunyaev-Zel'dovich effect (SZE) and X-ray
observations.  We model the intracluster medium (ICM) using a
spherical isothermal $\beta$ model.  We quantify the statistical and
systematic uncertainties inherent to these direct distance
measurements, and we determine constraints on the Hubble parameter for
three different cosmologies.  For an $\Om = 0.3$, $\Ol = 0.7$
cosmology, these distances imply a Hubble constant of $63
^{+12}_{-9\phn}\, ^{+21}_{-21}$ \ksM, where the uncertainties
correspond to statistical followed by systematic at 68\% confidence.
The best fit \Ho\ is 57 \ksM\ for an open $\Om = 0.3$ universe and 52
\ksM\ for a flat $\Om = 1$ universe.
\end{abstract}


\section{Introduction}
\label{sec:intro}
Analysis of Sunyaev-Zel'dovich effect (SZE) and X-ray data from a
cluster of galaxies provides information that can be used to determine
the distance to the cluster, independent of the extragalactic distance
ladder.  In the early seventies, Sunyaev and Zel'dovich (1970, 1972)
\nocite{sunyaev1970} \nocite{sunyaev1972} suggested that cosmic
microwave background (CMB) photons inverse Compton scattering off the
electrons in the hot ($\sim 10$ keV) intracluster medium (ICM) trapped
in the potential well of the cluster, would cause a small ($\lsim 1$
mK) distortion in the CMB spectrum, now known as the
Sunyaev-Zel'dovich effect (SZE).  The distortion appears as a
decrement for frequencies $\lsim 218$ GHz ($\lambda \gsim 1.4$ mm) and
as an increment for frequencies $\gsim 218$ GHz.  The SZE signal is
proportional to the pressure integrated along the line of sight
through the cluster, $\Delta T \sim \int\! n_e T_e d\ell$, where $n_e$
is the electron density of the ICM and $T_e$ is the electron
temperature.  The X-ray surface brightness can be written as $S_x \sim
\int\!  n_e^2 \LameH d\ell$ where \LameH\ is the X-ray cooling
function, which depends on temperature and metallicity.  It was soon
realized that one can determine the distance to the cluster by
capitalizing on the different dependencies on density, $n_e$, with
some assumptions about the geometry of the cluster.  This is a direct
distance based only on relatively simple cluster physics and not
requiring any standard candles or rulers.

The SZE signal is weak and difficult to detect.  The recent success of
SZE observations is due to advances in instrumentation and
observational strategy.  Recent high signal-to-noise ratio detections
have been made with single dish observations at radio wavelengths
\citep{birkinshaw1994, herbig1995, myers1997, hughes1998}, millimeter
wavelengths \citep{holzapfel1997b, holzapfel1997, pointecouteau1999}
and submillimeter wavelengths \citep{lamarre1998, komatsu1999}.
Interferometric observations have produced high quality images of the
SZE \citep{jones1993, grainge1993, carlstrom1996, carlstrom1998,
saunders1999, grainge1999}.  On the theoretical side, there is
substantial literature on relativistic corrections to both the SZE
\citep{rephaeli1997, itoh1998, challinor1998, nozawa1998b,
sazonov1998} and X-ray bremsstrahlung \citep{hughes1998, rephaeli1997,
nozawa1998}.  To date, there are about a dozen estimates of \Ho\ based
on combining X-ray and SZE data for individual clusters (see
\citealp{birkinshaw1999} for a comprehensive review).

\begin{deluxetable}{lcccccc}
\tablewidth{0pt}
\tablecaption{Radio Point Sources \label{tab:pt_sources}}
\tablehead{
  \colhead{} &
  \colhead{RA} & \colhead{DEC} & \colhead{$F_{28.5}$} &
  \colhead{$F_{15}$} & \colhead{$F_{5}$} & \colhead{$F_{1.4}$}\\
  \colhead{Field} &
  \colhead{(J2000)} & \colhead{(J2000)} & \colhead{(mJy)} &
  \colhead{(mJy)} & \colhead{(mJy)} & \colhead{(mJy)}
}
\tablecolumns{8}
\startdata
\cluster &
$04\h$ $54\m$ $22\s$ & $-03^\circ$ $01\arcmin$ $26\arcsec$ &
	1.88 & \nodata & \nodata & 14.9\tablenotemark{a} \\
\hline
\clustwo &
$00\h$ $18\m$ $31\s$ & $+16^\circ$ $20\arcmin$ $45\arcsec$ &
	9.07 & 25.0\tablenotemark{b}& 84.5\tablenotemark{b}
	& 267\tablenotemark{b} 
	
\enddata
\tablenotetext{a}{From Condon \etal\ (1998)\nocite{condon1998}.}
\tablenotetext{b}{From Moffet \& Birkinshaw (1989)\nocite{moffet1989}.}
\end{deluxetable}

We present a new analysis, wherein we perform a joint
maximum-likelihood fit to both interferometric SZE and X-ray data.
This method takes advantage of the unique properties of
interferometric SZE data, utilizing all the available image data on
the ICM.  This is the first time SZE and X-ray data have been analyzed
jointly.  We apply this method to observations of \clusterfull\ and
\clustwofull, massive clusters at redshift $z=0.55$
\citep{donahue1995, carlberg1994} and $z=0.5455$ \citep{neumann1997,
dressler1992}, respectively.

We describe the data and reduction in \S~\ref{sec:data}, the analysis
method in \S~\ref{sec:method}, and present the results and possible
systematic uncertainties in \S~\ref{sec:distances}.
Section~\ref{sec:disc_concl} contains a discussion of the results and
future prospects.  All uncertainties are 68.3\% confidence unless
explicitly stated otherwise.

\section{Observations and Data Reduction}
\label{sec:data}

\subsection{Interferometric SZE Observations}
\label{subsec:szobs}

\begin{figure*}[bth]
\epsfxsize = 7.5 in
\centerline{\epsfbox{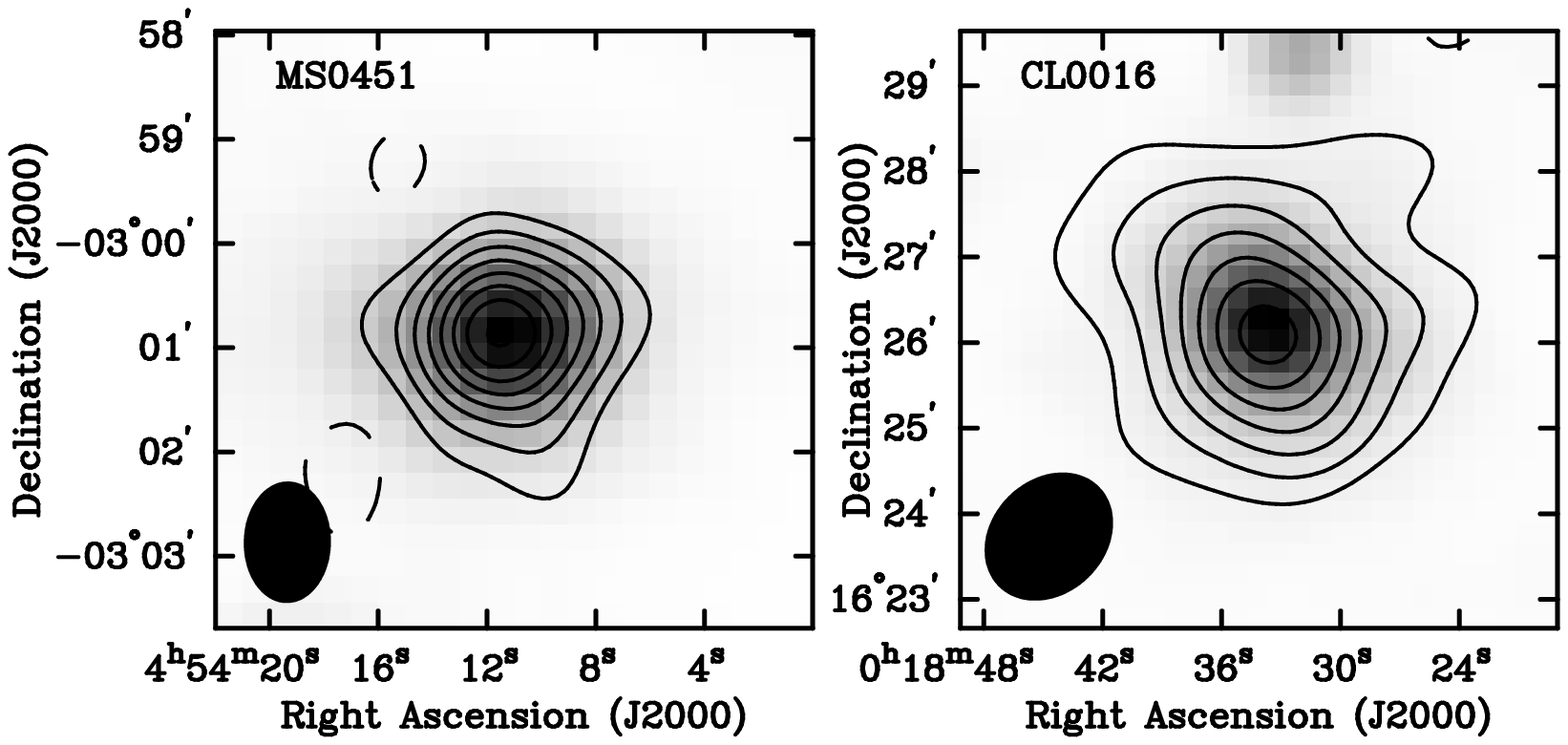}} 
\caption{SZE (contours) and X-ray (grey scale) image overlay for
         \clusterfull\ and \clustwofull.  The SZE images have an rms
         of $\sim 30$ $\mu$K (\cluster) and $\sim 46$ $\mu$K
         (\clustwo).  The contours are multiples of 2 $\sigma$ and
         negative contours are shown as solid lines.  The FWHM ellipse
         of the synthesized beam is shown in the lower left corner of
         each panel.  The different sizes of the SZE images are due to
         the smaller dishes of the BIMA telescopes (used for \clustwo)
         which can be more closely packed and thus sample smaller \uv\
         radii than the OVRO telescopes (used for \cluster).  The
         X-ray images are PSPC raw counts smoothed with a Gaussian
         with $\sigma = 15\arcsec$.  The peak for the \cluster\ image
         is 23 counts and 50 counts for the \clustwo\ image.
\label{fig:image}}
\end{figure*}

The extremely low systematics of interferometers and their
two-dimensional imaging capability make them well suited to study the
weak ($\lsim1$ mK) SZE signal in galaxy clusters.  Over the past
several summers, we outfitted the Berkeley Illinois Maryland
Association (BIMA) millimeter array in Hat Creek, California, and the
Owens Valley Radio Observatory (OVRO) millimeter array in Big Pine,
California, with centimeter wavelength receivers.  Our receivers use
cooled ($\sim 10$ K) High Electron Mobility Transistor (HEMT)
amplifiers \citep{pospieszalski1995} operating over 26-36 GHz with
characteristic receiver temperatures of $T_{rx}\sim $11-20 K, over
28-30 GHz, the band used for the observations presented here.  When
combined with the BIMA or OVRO systems, these receivers obtain typical
system temperatures scaled to above the atmosphere of $T_{sys}\sim 45$
K and as low as 34 K.  Most telescopes are placed close together in a
compact configuration to probe the angular scales subtended by distant
clusters ($\sim$ 1\arcmin), but telescopes are always placed at longer
baselines for simultaneous detection of point sources.  Every half
hour we observe a radio point source, commonly called a phase
calibrator, to monitor the system gains for about two minutes.

\cluster\ was observed at OVRO in 1996 during May and June for 30
hours with six 10.4 m telescopes using two 1 GHz channels centered at
28.5 GHz and 30.0 GHz (2 GHz bandwidth).  \clustwo\ was observed at
OVRO in 1994 between June 16 and July 4 for 87 hours with five 10.4 m
telescopes and a 1 GHz bandwidth centered at 28.7 GHz and in 1995
between July 24 and July 28 for 13 hours using five 10.4 m telescopes
and two 1 GHz channels centered at 28.5 GHz and 30.0 GHz.  \clustwo\
was also observed at BIMA in 1996 between September 6 and September 18
for 29 hours with six 6.1 m telescopes and in 1997 between June 21 and
July 22 for 8 hours with nine 6.1 m telescopes, both years with an 800
MHz bandwidth centered at 28.5 GHz.

The data are reduced using the MIRIAD \citep{sault1995} software
package at BIMA and using MMA \citep{scoville1993} at OVRO.  In both
cases, data are removed when one telescope shadows another, when
cluster data are not straddled by two phase calibrators, when there
are anomalous changes in instrumental response between calibrator
observations, or when there is spurious correlation.  For absolute
flux calibration, we use observations of Mars, on the assumption that
we know its true brightness temperature from the Rudy
(1987)\nocite{rudy1987} Mars model.  For observations not containing
Mars, calibrators in those fields are bootstrapped back to the nearest
Mars calibration (see \citealp{grego1999a} for more details).  The
observations of the phase calibrators over each summer give us a
summer-long calibration of the gains of the BIMA and OVRO
interferometers.  They both show very little gain variation, changing
by less than 1\% over a many-hour track, and the average gains remain
stable from day to day.

An interferometer samples the Fourier transform of the sky brightness
rather than the direct image of the sky.  The final products from the
interferometer are the amplitudes of the real and imaginary components
of the Fourier transform of the cluster SZE distribution on the sky
multiplied by the primary beam of the telescope.  The SZE data files
include the positions in the Fourier domain, which depend on the
arrangement of the telescopes in the array, the real and imaginary
components, and a measure of the noise in the real and imaginary
components.  The Fourier conjugate variables to right ascension and
declination are commonly called $u$ and $v$, respectively, and the
Fourier domain is commonly referred to as the \uv\ plane.

The finite size of each telescope dish imposes an almost Gaussian
attenuation across the field of view, known as the primary beam.  The
primary beams are constructed from holography data taken at each
array.  The main lobe of the primary beams can be approximated as
Gaussian with a full width at half maximum (FWHM) of $4\arcmin.2$ for
OVRO and $6\arcmin.6$ for BIMA at 28.5 GHz.  We use the primary beam
profiles made from the holography data for our analysis.

The primary beam sets the field of view.  The effective resolution,
called the synthesized beam, depends on the sampling of the \uv\ plane
of the observation and is therefore a function of the configuration of
the telescopes.  The cluster SZE signal is largest on the shortest
baselines (largest angular scales).  The shortest possible baseline is
set by the diameter of the telescopes, $D$.  Thus we are not sensitive
to angular scales larger than about $\lambda/2 D$, which is $\sim
2\arcmin.8$ for BIMA observations and $\sim 1\arcmin.7$ for OVRO
observations.  The compact configuration used for our observations
yields significant SZE signal at these angular scales, but the
interferometer is not sensitive to larger angular scales.  Because of
this spatial filtering by the interferometer, it is necessary to fit
models directly to the data in the \uv\ plane, rather than to the
deconvolved image.

Point sources are identified from SZE images created with DIFMAP
\citep{pearson1994} using only the long baseline data ($\gsim 2000\
\lambda$) and natural weighting.  Approximate positions and fluxes for
each point source are obtained from this image and used as inputs for
the model fitting discussed in \S~\ref{subsec:model_fit}.  The data
are separated by observatory, frequency, and by year to allow for
temporal and spectral variability of the point source flux.  One point
source is found in both the \cluster\ and \clustwo\ fields.  The point
source positions and fluxes are from the model fitting described in
\S~\ref{subsec:model_fit} and summarized in
Table~\ref{tab:pt_sources}.  Our positions agree very well with the
NRAO VLA Sky Survey (NVSS) source positions \citep{condon1998}.

The \cluster\ field point source is located 172\arcsec\ from the
pointing center with a measured flux of $0.50^{+0.07}_{-0.07}$ mJy at
28.5 GHz.  Correcting for the primary beam attenuation appropriate for
this offset, the intrinsic point source flux is $1.88^{+0.26}_{-0.26}$
mJy.  In the 30 GHz channel, this point source has a flux of
$0.36^{+0.07}_{-0.07}$ mJy, and $1.35^{+0.26}_{-0.26}$ after
correcting for the primary beam.  This point source was found in the
NVSS survey with a flux of 14.9 mJy at 1.4 GHz \citep{condon1998}.
The point source in the \clustwo\ field is 339\arcsec\ from the
pointing center and is only seen in the BIMA data since the OVRO
primary beam attenuation places it out of the OVRO field of view.  The
flux of this source is measured to be $1.01^{+0.23}_{-0.20}$ mJy at
28.5 GHz from the 1997 BIMA data, which when corrected for the primary
beam attenuation, is an intrinsic flux of $9.07^{+2.07}_{-1.80}$.
This source corresponds to source 15 from the survey of this field
done by Moffet \& Birkinshaw (1989)\nocite{moffet1989}.  They found
this source to be $267 \pm 3$ mJy at 1.44 GHz (264 mJy in the more
recent NVSS survey; \citealp{condon1998}), $84.5 \pm 1.1$ mJy at 4.86
GHz, and $25.0 \pm 1.5$ mJy at 14.94 GHz.  We do not see the other two
sources of Moffet \& Birkinshaw (1989)\nocite{moffet1989} within
360\arcsec\ of the pointing center, sources 10 and 14 which have 15
GHz fluxes of 0.56 and $<2.7$ mJy, respectively.  We do not present a
flux for the point source in the 1996 BIMA \clustwo\ data because of a
problem with the absolute calibration of the array during observations
early that summer.  Though the overall normalization is uncertain, the
data still provide shape information about the cluster.
Table~\ref{tab:pt_sources} summarizes the positions of the point
sources and their fluxes at various frequencies.

Figure~\ref{fig:image} shows the SZE image contours overlaid on the
X-ray images of these clusters.  We use DIFMAP \citep{pearson1994} to
produce the naturally weighted SZE images.  The point sources are
subtracted from the data and a Gaussian taper applied to emphasize
brightness variations on cluster scales before the image is
deconvolved (CLEANed).  For the \cluster\ OVRO data, we apply a 1200
$\lambda$ half-power radius Gaussian taper before deconvolving the
image.  This yields an elliptical Gaussian fit of $48\arcsec \times
70\arcsec$ for the synthesized beam (effective resolution) and a rms
of $\sim 68$ $\mu$Jy beam$^{-1}$, corresponding to a Rayleigh-Jeans
(RJ) brightness sensitivity of $\sim 30$ $\mu$K.  For \clustwo\ we use
the 1996 and 1997 BIMA data with a 1000 $\lambda$ half-power radius
Gaussian taper giving a $81\arcsec \times 101\arcsec$ synthesized beam
and a rms of $\sim 250$ $\mu$Jy beam$^{-1}$, corresponding to a $\sim
46$ $\mu$K RJ brightness sensitivity.  The SZE image contours are
multiples of twice the rms level for each image.  Images made with the
1994 OVRO \clustwo\ data were presented in Carlstrom, Joy, \& Grego
(1996)\nocite{carlstrom1996}.

We stress that these images are made to demonstrate the data quality.
The actual analysis is done in the Fourier plane, where the noise
characteristics of the data and the spatial filtering of the
interferometer are well understood.  The SZE and X-ray image overlays
in Figure~\ref{fig:image} show that the region of the cluster sampled
by the interferometric SZE observations and the X-ray observations is
similar.

\subsection{X-ray Observations}
\label{subsec:xobs}
We use archival {\it R\"{o}ntgen Satellite} ({\it ROSAT}) data from
both the Position Sensitive Proportional Counter (PSPC) and
High-Resolution Imager (HRI) instruments.  \cluster\ was observed with
the PSPC in 1993 over March 5-7 for 15,439 s of live-time and by the
HRI in 1995 over September 3-19 for 45,864 s of live-time.  There are
approximately 1200 cluster photons collected in both the PSPC and HRI
observations.  \clustwo\ was observed with the PSPC in 1992 over July
11-18 for a live-time of 41,589 s and by the HRI in 1995 between June
17 and July 5 for a live-time of 70,228 s.  The PSPC data contains
about 3200 cluster photons and the HRI has about 1500.

We use the Snowden Extended Source Analysis Software (ESAS)
\citep{snowden1994, snowden1998} to reduce the data.  We use this
software to generate a raw counts image, a noncosmic background image,
and an exposure map for the HRI (0.1-2.4 keV) data and for each of the
Snowden bands R4-R7 (PI channels $52-201$; approximately $0.5-2.0$
keV) for the PSPC data, using a master veto rate (a measure of the
cosmic-ray and $\gamma$-ray backgrounds) of 200 counts s$^{-1}$ for
the PSPC data.  We examine the light curve data of both instruments
looking for periodic, anomalously high count rates (short-term
enhancements) and for periods of high scattered solar X-ray
contamination.  None are found.  The Snowden software produces $512
\times 512$ pixel images with $14.947\arcsec$ pixels for the PSPC and
$5.0\arcsec$ pixels for the HRI.  For the PSPC, final images for all
of the R4-R7 bands together are generated by adding the raw counts
images and the background images.  Each Snowden band has a slightly
different effective exposure map and there is an energy dependence in
the point spread function (PSF).  Thus, we generate a single exposure
image and a single PSF image by combining cluster photon-weighted
averages of the four exposure images and the four PROS
\citep{worrall1992,conroy1993} generated on-axis PSF images.  The
cluster photon-weighting is determined using the background subtracted
detected photons within a circular region centered on the cluster.
The region selected to construct the weights is the largest circular
region encompassing the cluster which contains no bright point
sources.  For \cluster, we use a 12 pixel radius and for \clustwo\ we
use a 15 pixel radius.  X-ray images with SZE image overlays of
\cluster\ and \clustwo\ are shown in Figure~\ref{fig:image}.  The
gray-scale images are the PSPC ``raw'' counts images smoothed with a
Gaussian with $\sigma = 15\arcsec$.  The peaks of the images are 23
counts (\cluster) and 50 counts (\clustwo).

For \cluster, we use the emission-weighted temperature, galactic
absorption, and metallicity from Donahue (1996)\nocite{donahue1996}.
She found a best-fit X-ray temperature of $T_e = 10.4 ^{+1.0}_{-0.8}$
keV with a galactic absorption column density of $\NH = 3.0
^{+0.4}_{-0.3} \times 10^{20}$ cm$^{-2}$ and a metallicity of $0.15
^{+0.07}_{-0.07}$ solar implied by the iron abundance from a joint
analysis of {\it ASCA} and PSPC data.  This temperature is consistent
with the Mushotzky \& Scharf (1997)\nocite{mushotzky1997} value of
$T_e = 10.17 ^{+0.93}_{-0.76}$ keV.  For \clustwo, we adopt the Hughes
and Birkinshaw (1998)\nocite{hughes1998} results.  They found $T_e =
7.55 ^{+0.72}_{-0.58}$ keV with a galactic absorption column density
of $\NH = 5.59 ^{+0.41}_{-0.36} \times 10^{20}$ cm$^{-2}$ and a
metallicity of $0.07 ^{+0.11}_{-0.07}$ solar from a joint analysis of
{\it ASCA} and PSPC data.  This temperature agrees with a more recent
analysis by Furuzawa \etal\ (1998)\nocite{furuzawa1998} who found $T_e
= 8.0 ^{+0.6}_{-0.5}$ keV.  Unlike the Hughes \& Birkinshaw analysis,
this analysis did not include the PSPC data which is sensitive to the
column density.

\subsubsection{X-ray Cooling Function}
\label{subsec:emiss}
The Raymond-Smith (1977)\nocite{raymond1977} code calculates the
electron-ion bremsstrahlung contribution in the non-relativistic limit
using the Gaunt factors of Karzas \& Latter (1961)\nocite{karzas1961}.
Recently it has been pointed out \citep{rephaeli1997, hughes1998} that
relativistic corrections are important for a precise determination of
angular diameter distances, and therefore also of \Ho.  Though dubbed
``relativistic'' corrections, the corrections go beyond only
relativistic effects and include 1) relativistic corrections to the
electron distribution function, 2) relativistic and spin corrections
to the non-relativistic electron-ion bremsstrahlung cross section, 3)
inclusion of electron-electron bremsstrahlung (all three of relative
order $\kB T_e/m_ec^2$), and 4) first-order Born approximation
corrections to electron-ion bremsstrahlung (of order $(Ry/\kB
T_e)^{1/2}$, where $Ry=13.6$ eV is the ionization energy of hydrogen).
When applied, these corrections provide a better than 1\% accurate
calculation of thermal bremsstrahlung \citep{gould1980}.  Gould
provides results for both the integrated energy-loss rate and the
spectral cooling function from thermal bremsstrahlung.

Following Hughes \& Birkinshaw (1998), we use the corrections to the
spectral cooling function rather than the total energy-loss rate used
by Rephaeli \& Yankovitch (1997)\nocite{rephaeli1997} because the
calculated cooling function comes from integrating the spectral
cooling function over the fairly narrow {\it ROSAT} energy band
redshifted to the cluster frame.  These corrections may also affect
the $T_e$ derived from fits to X-ray spectral data because these
corrections modify the shape of the X-ray spectrum.  Hughes \&
Birkinshaw (1998)\nocite{hughes1998} found a $\sim 1$\% change in the
best fit $T_e$ for the Coma cluster ($T_e \sim 8$ keV) when applying
these corrections to their spectral fits.  As a check we also verified
that the spectral cooling function formula from Gould when integrated
over energy agrees with the total energy-loss result\footnote{We have
verified the misprint in Gould (1980)\nocite{gould1980} discussed by
Hughes \& Birkinshaw (1998)\nocite{hughes1998}.  This misprint
combined with the slightly different elemental compositions of the gas
considered in Rybicki \& Lightman (1979; RL)\nocite{rybicki1979} and
Rephaeli \& Yankovitch (1997) \nocite{rephaeli1997} explains the
difference between their total corrected energy-loss rates (RL pg.\
165).  The RL value is correct for the pure hydrogen gas they
consider.}.

\begin{deluxetable}{lccc}
\tablewidth{0pt}
\tablecaption{X-ray Cooling Functions for the PSPC\label{tab:emiss}}
\tablehead{
\colhead{} &
\colhead{\Lamo} &
\colhead{$\Lambda_{e \mbox{\tiny H} \circ}^{\mbox{\tiny
	det}}$} & 
\colhead{$\Sigma$}
\\
\colhead{Cluster} &
\colhead{(erg s$^{-1}$ cm$^3$)} &
\colhead{(cnt s$^{-1}$ cm$^5$)} &
\colhead{(\cgsunits)}
}
\tablecolumns{4}
\startdata
\cluster & $6.95 \times 10^{-24}$ & $3.26 \times 10^{-13}$ & $1.37 \times 10^{-11}$\\
\clustwo & $6.91 \times 10^{-24}$ & $3.00 \times 10^{-13}$ & $1.49 \times 10^{-11}$
\enddata
\end{deluxetable}

To calculate the X-ray spectral cooling function, we use a
Raymond-Smith (1993 Sep 21 version)\nocite{raymond1977} thermal plasma
model with its bremsstrahlung component replaced with Gould's
bremsstrahlung calculation including the corrections discussed above.
We use the Anders \& Grevesse (1989)\nocite{anders1989} meteoritic
abundances as the solar values, scaling the abundances of elements
heavier than He by the metallicity of the cluster.  We calculate the
absorption from cold Galactic gas using the photoelectric cross
sections from Balucinska-Church \& McCammon
(1992)\nocite{balucinska1992} (including the updated He absorption;
1993 Sep 23 version) for Anders \& Grevesse solar abundances.

We integrate the modified Raymond-Smith spectral model over the
redshifted {\it ROSAT} band (0.5-2.0 keV in the detector frame) to
determine the cooling function in cgs units, \Lamo.  We also calculate
the cooling function in detector units by multiplying the modified
Raymond-Smith spectrum by the response\footnote{Response matrices
obtained from ftp://legacy.gsfc.nasa.gov/caldb/data/, part of the
HEASARC calibration database.}  (includes effective area and energy
resolution) of the instrument, dividing by the energy of the photons
(to convert to counts), and integrating to find the total cooling
function, $\Lambda_{e \mbox{\tiny H}\circ}^{\mbox{\tiny det}}$.
Comparing these two yields the detector to cgs unit conversion,
$\Sigma$, after correcting by $(1+z)$ due to the difference between
instrument counts and energy ($\Sigma = \Lamo / \Lambda_{e \mbox{\tiny
H} \circ} ^{\mbox{\tiny det}} / (1+z)$).  The cooling function results
for the PSPC are summarized in Table~\ref{tab:emiss}.  The cooling
functions are 1.052 and 1.046 times the Raymond-Smith ``uncorrected''
value for \cluster\ and \clustwo, respectively.

\section{Method}
\label{sec:method}

\subsection{Angular Diameter Distance Calculation}
\label{subsec:da_calc}
The calculation begins by constructing a model for the cluster gas
distribution.  We use a spherical isothermal $\beta$ model to describe
the ICM.  Within this context the cluster's characteristic scale along
the line of sight is the same as the scale in the plane of the sky.
This model is clearly invalid in the presence of cluster
asphericities.  Thus cluster geometry introduces an important
uncertainty in SZE and X-ray derived distances.  In general, clusters
are dynamically young, are aspherical, and rarely exhibit projected
gas distributions which are circular on the sky \citep{mohr1995}.  We
currently can not disentangle the complicated cluster structure and
projection effects, but numerical simulations provide a good base for
understanding these difficulties.  The effects of asphericity
contribute significantly to the distance uncertainty for each cluster,
but do not result in any significant bias in the Hubble parameter
derived from a large sample of clusters \citep{sulkanen1999,
mohr1999b}.

The spherical isothermal $\beta$ model is given by
(\citealp{cavaliere1976}, 1978\nocite{cavaliere1978})
\begin{equation}
n_e({\mathbf{r}}) = \no \left ( 1 + \frac{r^2}{r_c^2} \right )^{-3\beta/2},
\label{eq:iso_beta}
\end{equation}
where $n_e$ is the electron number density, $r$ is the radius from the
center of the cluster, $r_c$ is the core radius of the ICM, and
$\beta$ is the power law index.  With this model, the SZE signal is
\begin{eqnarray}
\Delta T &=& f_{(x)} \Tcmb \Da \! \int\!\! d\zeta \, \sigT n_e \frac{\kB
	T_e}{m_e c^2} \nonumber\\
	 &=& \dTo \left ( 1 +
	\frac{\theta^2}{\theta_c^2} \right )^{(1-3\beta)/2},
	\label{eq:szsignal}
\end{eqnarray}
where $\Delta T$ is the SZE decrement/increment, $f_{(x)}=\fx$
($f_{(x)}\rightarrow -2$ in the non-relativistic and Rayleigh-Jeans
limits) is the frequency dependence of the SZE with $x = h\nu/k\Tcmb$,
$\delta_{\mbox{\tiny SZE}}(x,T_e)$ is the relativistic correction to
the frequency dependence, $\Tcmb$ (=2.728 K; \citealp{fixsen1996}) is
the temperature of the CMB radiation, \kB\ is the Boltzmann constant,
$\sigT$ is the Thompson cross section, $m_e$ is the mass of the
electron, $c$ is the speed of light, \dTo\ is the central SZE
decrement/increment, $\theta$ is the angular radius in the plane of
the sky and $\theta_c$ the corresponding angular core radius, and the
integration is along the line of sight $\ell=\Da\zeta$.  We apply the
relativistic corrections $\delta_{\mbox{\tiny SZE}}$ to fifth order in
$kT_e/m_e c^2$ \citep{itoh1998}.  The Itoh \etal\ results agree with
other work \citep{stebbins1997, challinor1998} to third order where
they stop.  This correction decreases the magnitude of $f_{(x)}$ by
3.7\% for \cluster\ and 2.7\% for \clustwo.  The correction is
slightly higher for \cluster, as expected, because of its higher
electron temperature.

The X-ray surface brightness is 
\begin{eqnarray}
S_x &=& \frac{1}{4\pi (1+z)^4} \Da \! \int \!\! d\zeta \, n_e \nH \LameH
	\:\;\! \nonumber\\
    &=& \Xo \left ( 1 + \frac{\theta^2}{\theta_c^2} \right
	)^{(1-6\beta)/2},  \label{eq:xsignal}
\end{eqnarray}
where $S_x$ is the X-ray surface brightness in cgs units (erg s$^{-1}$
cm$^{-2}$ arcmin$^{-2}$), $z$ is the redshift of the cluster, \nH\ is
the hydrogen number density of the ICM, $\LameH = \LameH (T_e,
\mbox{abundance})$ is the X-ray cooling function of the ICM in the
cluster rest frame in cgs units (erg cm$^3$ s$^{-1}$) integrated over
the redshifted {\it ROSAT} band, and \Xo\ is the X-ray surface
brightness in cgs units at the center of the cluster.  Since the X-ray
observations are in instrument counts, we also need the conversion
factor between detector counts and cgs units, $\Sigma$, discussed in
\S\ref{subsec:emiss} ($\Xo = S_{x \circ}^{\mbox{\tiny det}} \Sigma$).
The normalizations, \dTo\ and \Xo, used in the fit include all of the
physical parameters and geometric terms that come from the integration
of the $\beta$ model along the line of sight.

One can solve for the angular diameter distance by eliminating \no\
(noting that $\nH = n_e \mu_e / \muH$ where $n_j \equiv \rho/\mu_j
m_p$ for species $j$) yielding
\begin{eqnarray}
\Da = \frac{(\dTo)^2}{\Xo} \left( \frac{m_e c^2}{\kB \Teo} \right)^2
	\frac{\Lamo \mu_e / \muH}{4\pi^{3/2} \, f_{(x)}^2 \,
	T_{\mbox{\tiny CMB}}^2 \, \sigma_{\mbox{\tiny T}}^2 \,
	(1+z)^4} \nonumber \\
\times
	\frac{1}{\theta_c} \left [
	\frac{\Gamma(\frac{3}{2}\beta)}{\Gamma(\frac{3}{2}\beta-\frac{1}{2})}
	\right ]^2 \frac{\Gamma(3\beta-\frac{1}{2})}{\Gamma(3\beta)}
	\label{eq:Da}
\end{eqnarray}
where $\Gamma(x)$ is the Gamma function.  Similarly, one can eliminate
$\Da$ instead and solve for the central density \no.

\subsection{Joint SZE and X-ray Model Fitting}
\label{subsec:model_fit}
The SZE and X-ray emission both depend on the properties of the ICM,
so a joint fit to all the available data provides the best constraints
on those properties.  We perform a joint fit to the interferometric
SZE data and the PSPC and HRI X-ray data.  Each data set is assigned a
collection of parameterized models.  Typically, SZE data sets are
assigned a $\beta$ model and point sources and X-ray images are
assigned a $\beta$ model and a cosmic X-ray background model.  This
set of models is combined for each data set to create a composite
model which is then compared to the data.

Model parameters can be fixed, free to find their optimized values, or
gridded.  They can also be linked, forced to vary together among the
data sets.  In practice, $\theta_c$ and $\beta$ are linked between all
data sets (both SZE and X-ray) and the central decrements \dTo\ are
linked between the SZE data sets which are separated by season and
array.  We use a downhill simplex to search parameter space and
maximize the joint likelihood \citep{press1992}.  The cluster
position, $\beta$, $\theta_c$, \Xo, \dTo, a constant cosmic
background, radio point source positions, and point source fluxes are
all allowed to vary.

Each data set is independent, and likelihoods from each data set can
simply be multiplied together to construct the joint likelihood.
Likelihood ratio tests can then be performed to get confidence regions
or compare two models.  Rather than working directly with likelihoods,
$\mathcal{L}$, we work with $S \equiv -2\ln(\mathcal{L})$.  We then
construct a $\Delta \chi^2$-like statistic from the log likelihoods,
$\Delta S \equiv S_n - S_{ref}$ where $S_{ref}$ is the reference $S$
statistic, typically chosen to be the minimum of the $S$ function, and
$S_n$ is the $S$ statistic where $n$ parameters differ from the
parameters at the reference.  The statistic $\Delta S$ is sometimes
referred to as the Cash (1979)\nocite{cash1979} statistic and tends to
a $\chi^2$ distribution with $n$ degrees of freedom
(\citealp{kendall1979} for example).  This $\Delta S$ statistic is
equivalent to the likelihood ratio test and is used to generate
confidence regions and confidence intervals with $S_{ref}=S_{min}$.
For one interesting parameter, the 68.3\% ($\sim 1\sigma$) confidence
level corresponds to $\Delta S = 1.0$.

Because we are interested only in differences in log likelihoods,
$\Delta S$, the model independent terms in the likelihoods are
dropped.  The log likelihoods are then
\begin{eqnarray}
\sum_i -\frac{1}{2} \left ( \Delta R^2_i + \Delta I^2_i \right )
	W_i & & \mbox{for SZE data (Gaussian)}, \label{eq:likesz} \\
\sum_i D_i\ln(M_i) - M_i \ \ \ \ \ \ \: \,
	& & \mbox{for X-ray data (Poisson)}, \label{eq:likex}
\end{eqnarray}
where $\Delta R_i$ and $\Delta I_i$ are the differences between the
model and data at each point $i$ in the Fourier plane for the real and
imaginary components respectively, $W_i = 1/\sigma_i^2$ is a measure
of the noise (Gaussian) of the real and imaginary components discussed
in \S\ref{subsec:szobs}, and $M_i$ and $D_i$ are the model prediction
and data in pixel $i$.

\begin{figure*}[htb]
\hbox{
  \epsfxsize = 3.5 in
  \epsfbox{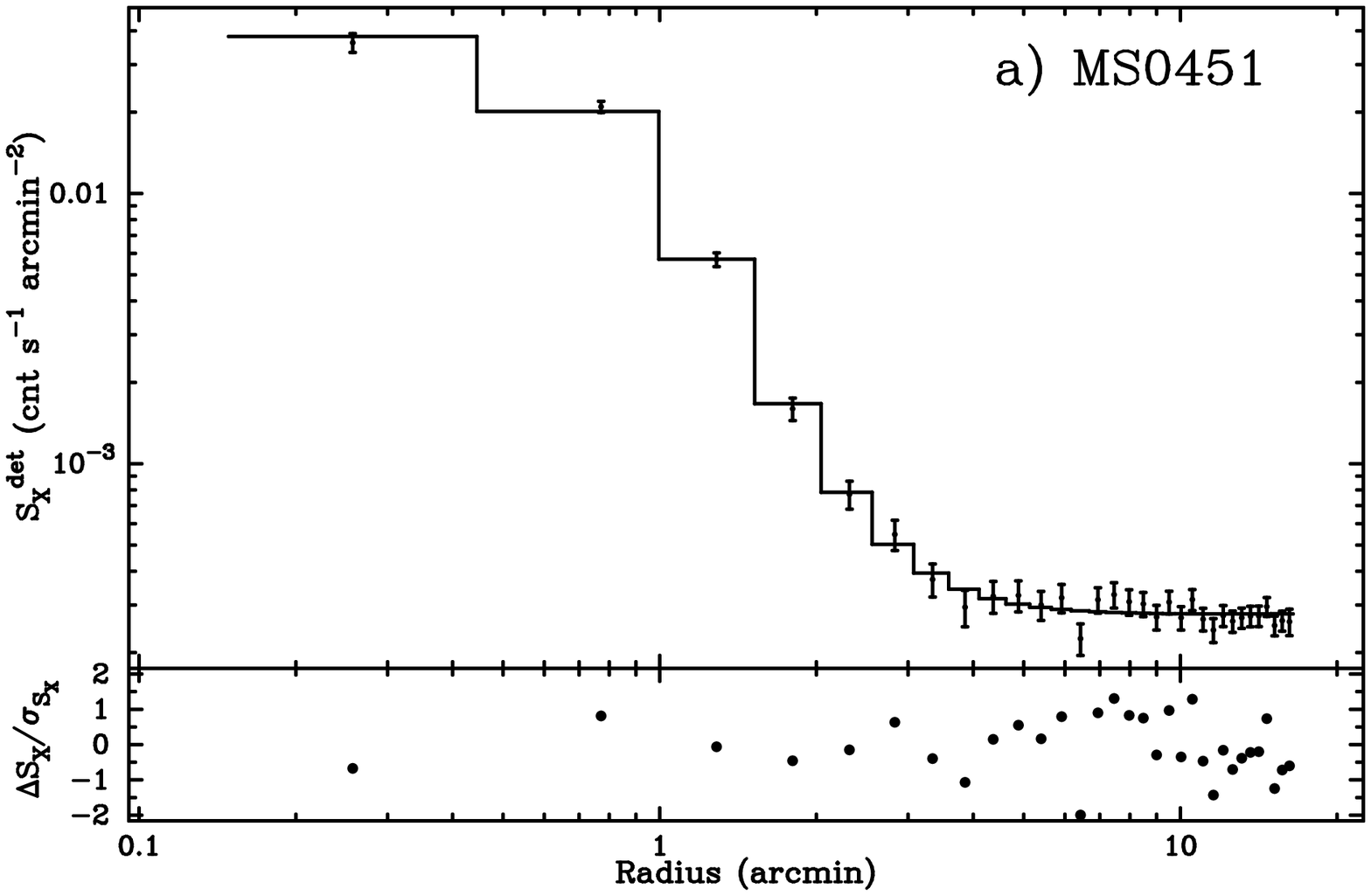}
  \epsfxsize = 3.5 in
  \epsfbox{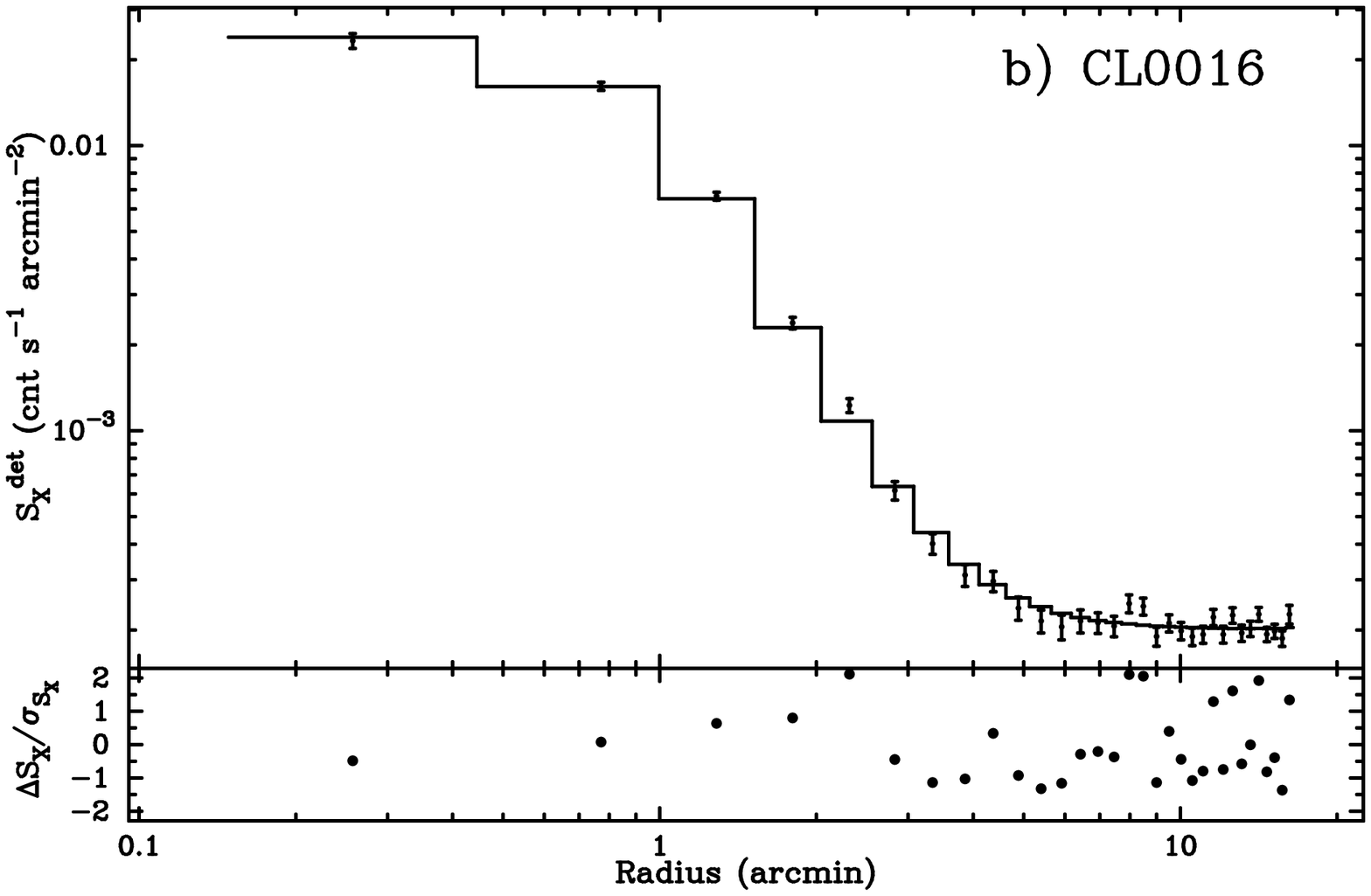}
} 
\caption{Comparison of the PSPC radially averaged surface brightness
  	 profile (points with error bars) with the best fitting
  	 isothermal spherical $\beta$-model plus background
  	 (histogram) for a) \cluster\ and b) \clustwo.  The lower
  	 panel shows the residuals in units of the standard deviation.
  	 The best fit model is a good fit to the data over the entire
  	 range of radii considered in the fit.  There is no evidence
  	 of excess emission near the core for either cluster, the
  	 signature of cooling flows.
\label{fig:radprof}}
\end{figure*}

The interferometric SZE observations provide constraints in the
Fourier (\uv) plane, so we perform our model fitting in the \uv\
plane, where the noise properties of the data and the spatial
filtering of the interferometer are well defined.  The SZE composite
model for both \cluster\ and \clustwo\ consists of a $\beta$ model and
a point source.  The $\beta$ model is computed in a regular grid in
image space, multiplied by the primary beam determined from holography
measurements, and fast Fourier transformed to produce the \uv\ plane
model.  It is then interpolated to the \uv\ position for each data
point.  Point sources are computed analytically at each \uv\ data
point of the observation and added to the $\beta$ model in the \uv\
plane to construct the composite SZE model.  The Gaussian likelihood
(eq.~[\ref{eq:likesz}]) is calculated using the composite SZE model
and the SZE data.  During the fitting, the cluster center, $\theta_c$,
$\beta$, \dTo, the point source positions, and the point source fluxes
are all allowed to vary.

Because the SZE is frequency dependent, a minor additional detail
comes from observations of a cluster at multiple frequencies, 28.5 GHz
and 30 GHz.  We input a central decrement appropriate for 30 GHz into
the fitting routine, which then corrects the model for the actual
frequency of the observation. The likelihood is calculated with the
model appropriate for the observing frequency.  This allows us to link
the central decrement across data sets with different observing
frequencies.

\begin{deluxetable}{lccccc}
\tablewidth{0pt}
\tablecaption{ICM Parameters \label{tab:fitparam}}
\tablehead{
\colhead{} & 
\colhead{} & 
\colhead{$\theta_c$} & 
\colhead{$S_{x\circ}^{\mbox{\tiny det}}$} & 
\colhead{\Xo} & 
\colhead{\dTo} \\
\colhead{Cluster} & 
\colhead{$\beta$} & 
\colhead{(arcsec)} & 
\colhead{(cnt s$^{-1}$ arcmin$^{-2}$)} & 
\colhead{(erg s$^{-1}$ cm$^{-2}$ arcmin$^{-2}$)} & 
\colhead{($\mu$K)}
}
\tablecolumns{6}
\startdata
\cluster & $0.806^{+0.052}_{-0.043}$ & $34.7^{+3.9}_{-3.5}$ &
	$6.96^{+0.63}_{-0.61}$ $\times 10^{-2}$ & 
	$9.56^{+0.86}_{-0.84}$ $\times 10^{-13}$ &
	$-1431^{+\phn 98}_{-\phn 93}$ \\
\clustwo & $0.749^{+0.024}_{-0.018}$ & $42.3^{+2.4}_{-2.0}$ &
	$4.14^{+0.16}_{-0.19}$ $\times 10^{-2}$ & 
	$6.17^{+0.24}_{-0.28}$ $\times 10^{-13}$ &
	$-1242^{+105}_{-105}$
\enddata
\end{deluxetable}

The model for each X-ray data image includes a spherical isothermal
$\beta$ model plus a constant cosmic background.  The model is
convolved with the appropriate PSF, multiplied with the exposure map,
and then the noncosmic background is added pixel by pixel.  Point
sources are masked out.  The logarithm of the Poisson likelihood
(eq.~[\ref{eq:likex}]) is then calculated.  During the fitting, the
cluster center, $\theta_c$, $\beta$, \Xo, and the cosmic background
are all allowed to vary.  The PSF is generated by PROS and the
exposure map and noncosmic background maps are those generated by the
Snowden ESAS software discussed in \S~\ref{subsec:xobs}.  The PSF and
exposure maps for the PSPC Snowden bands R4-R7 are combined in a
cluster photon-weighted average (see \S~\ref{subsec:xobs}).  Point
sources are found using the ESAS detection algorithm with a 3 $\sigma$
detection criterion and masked out.  Circular regions of typically 3
pixel radius are placed on each point source and excluded from the
calculation of the likelihood.  These regions correspond to radii of
$\sim 45\arcsec$ for the PSPC and $15\arcsec$ for the HRI.  As a
check, the image of the cluster excluding the masked regions is
visually inspected.  Increasing the size of the masked regions does
not significantly alter the best fit parameters, including the cosmic
background.  For the model fitting we use a region centered on the
cluster with a 64 pixel radius, corresponding to $\sim 16\arcmin$ for
the PSPC and $\sim 5\arcmin$ for the HRI.  Using a larger fitting
region does not change the best fit model parameters significantly.

When allowed to vary separately, the best-fit central surface
brightnesses for the PSPC and HRI are consistent within their
uncertainties when compared in cgs units.  We linked the central
surface brightnesses between the PSPC and HRI in cgs units, using
$\Sigma$ to convert to counts before comparing with the X-ray images.
The linked \Xo\ case gives an insignificant change in the $S$
statistic compared to the case where the PSPC and HRI normalizations
are allowed to vary individually and removes one free parameter.  The
consistent central surface brightnesses from the PSPC and HRI
observations present an interesting test of the relative calibration
of the two instruments.

\section{Direct Distances and the Hubble Constant}
\label{sec:distances}
The results from our maximum likelihood joint fit to the SZE and X-ray
data are summarized in Table~\ref{tab:fitparam} for both
clusters.
Figures~\ref{fig:radprof}{\it a} and 2{\it b} show the X-ray radial
surface brightness profiles and the best fit composite models for
\cluster\ and \clustwo, respectively.  The models for both clusters
show a good fit to the data over a large range of angular radii.
Using equation~(\ref{eq:Da}) with the best-fit parameters from
Table~\ref{tab:fitparam} and the cooling functions from
Table~\ref{tab:emiss}, we find the distance to \cluster\ to be
$1278^{+265}_{-298}$ Mpc and the distance to \clustwo\ to be
$2041^{+484}_{-514}$ Mpc, where the uncertainties are statistical only
(see discussion below).

Our fitting results are consistent with previous analyses of the {\it
ROSAT} data of \cluster\ and \clustwo.  Donahue
(1996)\nocite{donahue1996} analyzed the \cluster\ PSPC data and found
$\beta = 1.01^{+0.27}_{-0.18}$ and $\theta_c = 38.2^{+11.6}_{-\phn
9.6}$ arcseconds.  Table~\ref{tab:cl0016} shows the comparison for
\clustwo\ with Neumann \& B\"{o}hringer (1997)\nocite{neumann1997} and
Hughes \& Birkinshaw (1998)\nocite{hughes1998}.

\begin{deluxetable}{lcll}
\tablewidth{0pt}
\tablecaption{Comparison of \clustwo\ Analyses\label{tab:cl0016}}
\tablehead{
& & & \colhead{$\theta_c$}\\
\colhead{Reference}  & 
\colhead{Instrument} &
\colhead{$\beta$} &
\colhead{(arcsec)}
}
\tablecolumns{4}
\startdata
NB97\tablenotemark{a} & PSPC 		  & $0.80\phn^{+0.04}_{-0.05}$    
	& $50.5^{+4.5}_{-4.0}$\\
NB97\tablenotemark{a} & HRI  		  & $0.68\phn^{+0.10}_{-0.07}$    
	& $38.5^{+8.0}_{-6.5}$\\
HB98\tablenotemark{b} & PSPC 		  & $0.728^{+0.025}_{-0.022}$ 
	& $40.7^{+2.7}_{-2.3}$\\
This work             & joint PSPC \& HRI & $0.749^{+0.024}_{-0.018}$ 
	& $42.3^{+2.4}_{-2.0}$
\enddata
\tablenotetext{a}{\citealp{neumann1997}}
\tablenotetext{b}{\citealp{hughes1998}}
\end{deluxetable}

We also compare our results with the distance determination to
\clustwo\ by Hughes \& Birkinshaw (1998)\nocite{hughes1998}.  They
analyzed the PSPC observations to determine the ICM shape parameters
and then used that model to extract the SZE central decrement from
observations taken with the OVRO single dish 40~m telescope at 20.3
GHz.  They observed seven points in a north-south scan through
\clustwo.  Beam switching was done using the 40~m dual-beam system,
which provides two 1\arcmin.78 FWHM beams separated by 7\arcmin.15 in
azimuth.  The central decrement extracted from such scans depends on
the adopted center of the SZE signal as well as the adopted ICM shape
parameters, $\beta$ and $\theta_c$.  Interferometric observations
provide two-dimensional imaging information with accurate astrometry
and therefore provide information about the cluster center and the ICM
shape parameters.  We find a central decrement of
$-1242^{+105}_{-105}$ $\mu$K remarkably consistent with the Hughes \&
Birkinshaw value of $-1201^{+189}_{-189}$ $\mu$K converted to
thermodynamic temperature at 30 GHz.  Hughes \& Birkinshaw found the
distance to \clustwo\ to be $1863 ^{+836} _{-549}$ Mpc in good
agreement with ours, where the uncertainty is statistical only and we
have corrected for the frequency dependence of the SZE ($f_{(x)}\neq
-2$) and relativistic corrections.  Our PSPC central surface
brightness and cooling function are both lower than the Hughes \&
Birkinshaw values.  This difference arises entirely from using a
different bandpass for the analysis ($0.5-2.0$ keV versus $0.4-2.4$
keV).  However, only the ratio of the surface brightness and cooling
function enters into the distance calculation.  Our ratio times
$\mu_H/\mu_e$ (\LameH\ versus \Lamee) is $1.65 \times 10^{11}$
arcmin$^{-2}$ cm$^{-5}$ in fortuitously good agreement with theirs,
$1.64 \times 10^{11}$ arcmin$^{-2}$ cm$^{-5}$.

\begin{deluxetable}{lccccc}
\tablewidth{0pt}
\tablecaption{\Da\ Observational Uncertainty Budget (percent)
	\label{tab:Daerr}}
\tablehead{
\colhead{Cluster} & 
\colhead{Fit\tablenotemark{a}} & 
\colhead{$\NH$\tablenotemark{b}} & 
\colhead{$[\mbox{Fe}]/[\mbox{H}]$\tablenotemark{c}} &
\colhead{$T_e$\tablenotemark{b}} & 
\colhead{Total\tablenotemark{d}}
}
\tablecolumns{6}
\startdata
\cluster & $^{+13.7}_{-13.1}$ & $^{+0.9}_{-1.2}$ &  
	$^{+1.1}_{-1.1}$ & $^{+15.4}_{-19.2}$ &
	$^{+20.7}_{-23.3}$\\
\clustwo & $^{+17.8}_{-16.4}$ & $^{+1.1}_{-1.2}$ & 
	$^{+2.1}_{-1.3}$ & $^{+15.4}_{-19.1}$ &
	$^{+23.7}_{-25.2}$
\enddata
\tablenotetext{a}{The 68.3\% uncertainties over the
	four-dimensional error surface for $\beta$, $\theta_c$, \Xo,
	and \dTo.}
\tablenotetext{b}{\Da\ decreases as parameter increases.}
\tablenotetext{c}{Metallicity relative to solar.}
\tablenotetext{d}{Combined in quadrature.}
\end{deluxetable}

There is a known correlation between the $\beta$ and $\theta_c$
parameters of the $\beta$ model.  One might think this correlation
would make determinations of \Da\ imprecise because \Da\ is calculated
from these very shape parameters of the ICM.  Figure~\ref{fig:chisq}
illustrates this correlation and its effect on \Da\ for \cluster.  The
filled contours are the 1, 2, and 3 $\sigma$ $\Delta S$ confidence
regions for $\beta$ and $\theta_c$ jointly.  The lines are contours of
constant \Da\ in Mpc.  With our interferometric SZE data, the contours
of constant \Da\ lie roughly parallel
\myputfigure{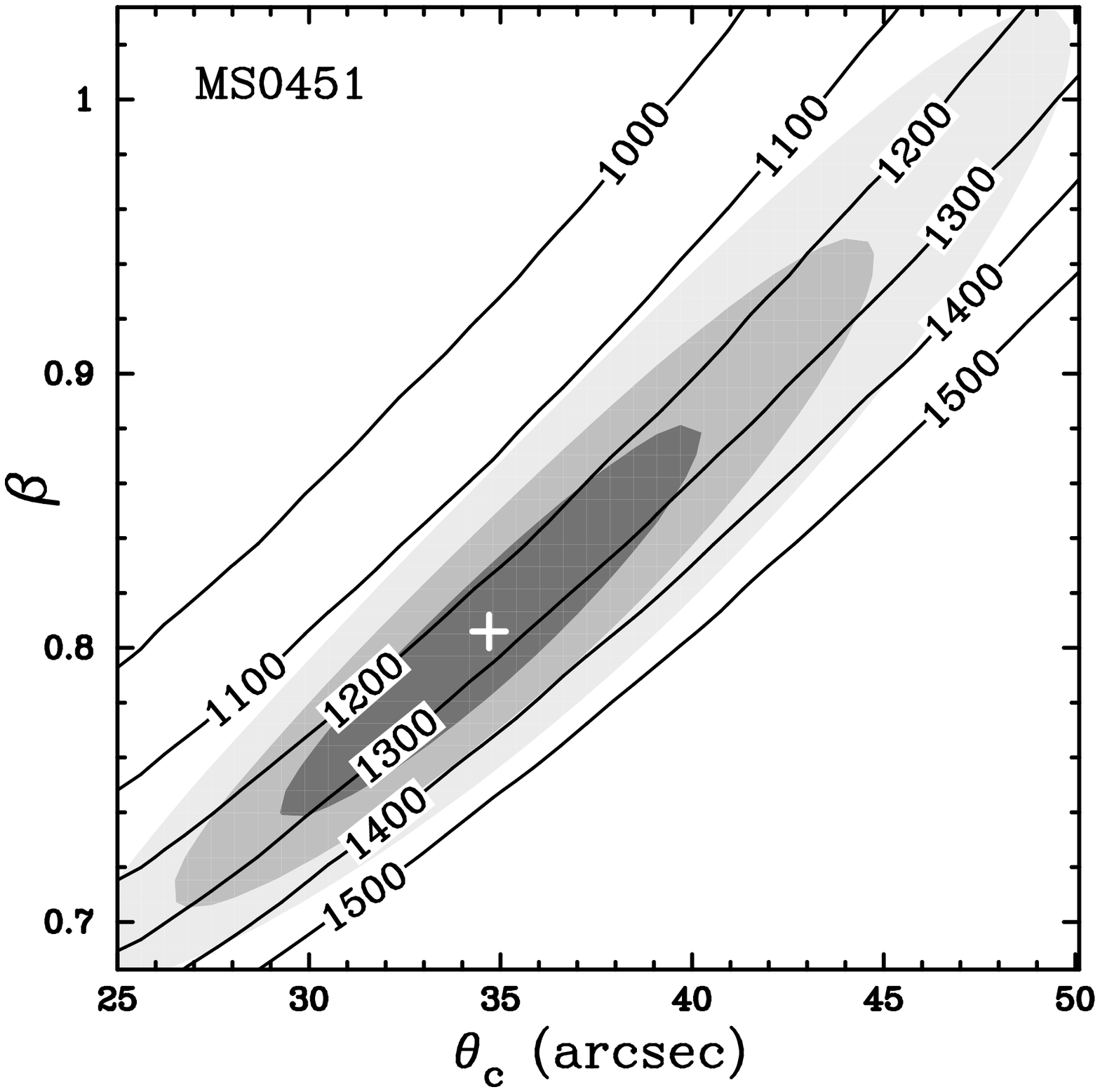}{2.8}{0.45}{-15}{5}
\figcaption{Confidence regions from the joint SZE and X-ray fit for
	   \cluster.  The filled regions are 1, 2, and 3 sigma
	   confidence regions for $\beta$ and $\theta_c$ jointly
	   ($\Delta S=$ 2.3, 6.2, 11.8) and the cross marks the best
	   fit $\beta$ and $\theta_c$. Solid lines are contours of
	   angular diameter distance in Mpc.  The \Da\ contours lie
	   roughly parallel to the $\beta$-$\theta_c$ correlation,
	   minimizing the effect of this correlation on the
	   uncertainties of \Da.
\label{fig:chisq}}
\medskip
\noindent to the $\beta$-$\theta_c$
correlation, minimizing the effect of this correlation on the
uncertainties of \Da.  Figure~\ref{fig:chisq2} shows similar behavior
for \clustwo.  Different observing techniques will result in different
behavior.  Contours of constant \Da\ have been found to be roughly
orthogonal to the $\beta$-$\theta_c$ correlation for some single dish
SZE observations \citep{birkinshaw1994,birkinshaw1991}.

Uncertainties in the angular diameter distance from the fit parameters
are calculated by gridding in the interesting parameters to explore
the $\Delta S$ likelihood space.  The most important parameters in
this calculation are \dTo, \Xo, $\beta$, and $\theta_c$.  Radio point
sources and the cosmic X-ray background affect \dTo\ and \Xo,
respectively.  Therefore we grid in \dTo, \Xo, $\beta$, and $\theta_c$
allowing the X-ray backgrounds for the PSPC and HRI to float
independently.  To estimate the effect of the radio point sources, we
find the best-fit parameter values with the point source flux fixed at
its best fit value.  We also run the grids for the point source flux
fixed at the $\pm 1\ \sigma$ values.  The point sources in the
\cluster\ and \clustwo\ fields are both far enough from the cluster
center so that their flux contributions have \, a 
\myputfigure{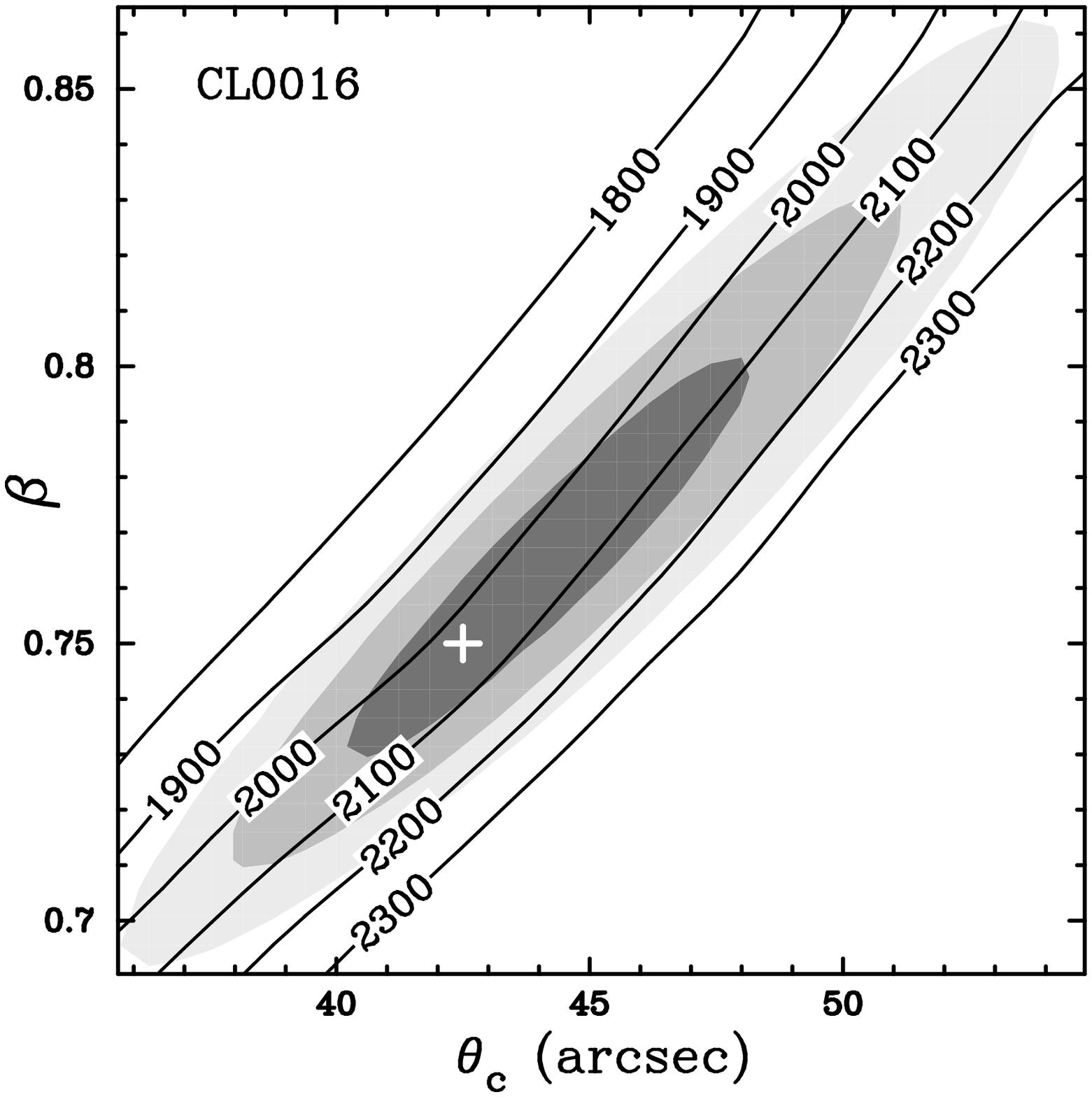}{2.8}{0.45}{-15}{5}
\figcaption{Same as Figure~\ref{fig:chisq} for \clustwo.
\label{fig:chisq2}}\medskip
\noindent negligible
effect on the central decrement and do not change the cluster shape
parameters significantly.  From this four dimensional $\Delta S$
hyper-surface, we construct confidence intervals for each parameter
individually as well as confidence intervals for \Da\ due to \Xo,
\dTo, $\beta$, and $\theta_c$ jointly.  The correlations between the
$\beta$ model parameters require this treatment to determine
accurately the uncertainty in \Da\ from the fitted parameters.  To
compute the 68.3\% confidence region we find the minimum and maximum
values of the parameter within a $\Delta S$ of 1.0.  We emphasize that
these uncertainties are meaningful only within the context of the
spherical isothermal $\beta$ model.

\begin{deluxetable}{lc}
\tablewidth{0pt}
\tablecaption{\Ho\ Systematic Uncertainty Budget for \cluster\ and \clustwo
	\label{tab:Hosyserr}}
\tablehead{
\colhead{Systematic} 	& \colhead{Effect}\\
& \colhead{(\%)}
}
\tablecolumns{2}
\startdata
SZE calibration			&$\pm \phn 8$\\
X-ray calibration		&$\pm 10$\\
\NH				&$\pm \phn 5$\\
Asphericity\tablenotemark{a}	&$\pm 14$\\
Isothermality \& clumping	&$\pm 20$\\
Undetected radio sources\tablenotemark{b}
				&$\pm 16$\\
Kinetic SZE\tablenotemark{a}	&$\pm \phn 6$\\
\hline
Total\tablenotemark{c}		&$\pm 33$
\enddata
\tablenotetext{a}{Includes a $1/\sqrt{2}$ factor for our 2 cluster sample.}
\tablenotetext{b}{Average of effect from the two cluster fields.}
\tablenotetext{c}{Combined in quadrature.}
\end{deluxetable}

The observational uncertainty budget for \Da\ is shown in
Table~\ref{tab:Daerr}.  The uncertainties in the fitted parameters
come from the above procedure.  The only other parameter that enters
directly into the \Da\ calculation is \Teo.  Since $\Da \propto
\Teo^{-2}$, the uncertainty in \Da\ due to \Teo\ is listed as twice
the fractional uncertainty on \Teo.  The other parameters, column
density and metallicity, as well as \Teo, affect the X-ray cooling
function.  We estimate the uncertainties in \Da\ due to these
parameters by taking their 68.3\% ranges and seeing how much they
affect the cooling function.  The uncertainty in the cooling function
due to \Teo\ is $\lsim 0.5$\% and is ignored.  The uncertainty on \Da\
due to observations is dominated by the uncertainty in the electron
temperature and the SZE central decrement.  Note that changes of
factors of two in metallicity result in a $\sim 1$\% effect on \Da.
The column densities measured from the X-ray spectra are different
from those from \ion{H}{1} surveys \citep{dickey1990}.  We use the
column densities from X-ray spectral fits since that includes
contributions from nonneutral hydrogen and other elements which
absorb X-rays.  The survey derived column densities change the angular
diameter distance by $\sim \pm 5$\%, which we include as a systematic
uncertainty (see \S~\ref{sec:disc_concl}).

To determine the Hubble Constant, we perform a $\chi^2$ fit to our
calculated \Da's versus $z$ for three different cosmologies.  To
estimate statistical uncertainties, we combine the uncertainties on
\Da\ listed in Table~\ref{tab:Daerr} in quadrature, which is only
strictly valid for Gaussian distributions.  This combined statistical
uncertainty is symmetrized (averaged) and used in the fit.  We find
\begin{equation}
\Ho = \cases{
	52 ^{+10}_{-\phn 7} \ \ksM; &\Om=1.0, \Ol=0.0, \cr
	57 ^{+11}_{-\phn 8} \ \ksM; &\Om=0.3, \Ol=0.0, \cr
	63 ^{+12}_{-\phn 9} \ \ksM; &\Om=0.3, \Ol=0.7, \cr
}
\label{eq:Horesult}
\end{equation}
where the uncertainties are statistical only.  The statistical error
comes from the $\chi^2$ analysis and includes uncertainties from
$T_e$, the parameter fitting, metallicity, and \NH\ (see
Table~\ref{tab:Daerr}).  We have chosen three cosmologies encompassing
the currently favored models.  There is a $\sim 20$\% range in \Ho\ at
$z \sim 0.5$ due to the geometry of the universe.

\subsection{Sources of Possible Systematic Uncertainty}
\label{subsec:systematic}
The absolute calibration of both the SZE observations and the PSPC and
HRI directly affects the distance determinations.  The absolute
calibration of the interferometric observations is conservatively
known to about 4\% at 68\% confidence, corresponding to a 8\%
uncertainty in \Ho\ ($\propto \Delta T_\circ^{-2}$).  The effective
areas of the PSPC and HRI are thought to be known to about 10\%,
introducing a 10\% uncertainty into the \Ho\ determination through the
calculation of $\Sigma$.  In addition to the absolute calibration
uncertainty from the observations, there are possible sources of
systematic uncertainty that depend on the physical state of the ICM
and other sources that can contaminate the cluster SZE emission.
Table~\ref{tab:Hosyserr} summarizes the systematic uncertainties in
the Hubble constant determined from \cluster\ and \clustwo.

\subsubsection{Cluster Atmospheres and Morphology}
\label{subsubsec:atmos_morph}
Most clusters do not appear circular in radio, X-rays, or optical.
Fitting a projected elliptical isothermal $\beta$ model gives an axial
ratio of $\sim 0.80$ and $\sim 0.84$ for \cluster\ and \clustwo,
respectively, close to the local average of $0.80$ \citep{mohr1995}.
Under the assumption of axisymetric clusters, the combined effect of
cluster asphericity and its orientation on the sky conspires to
introduce a $\sim\pm 20$\% random uncertainty in \Ho\
\citep{hughes1998}.  When one considers a large, unbiased sample of
clusters, presumably with random orientations, the uncertainty due to
imposing a spherical model will cancel, manifesting itself in the
statistical uncertainty and allowing a precise determination of \Ho.
Recently, Sulkanen (1999)\nocite{sulkanen1999} studied projection
effects using triaxial $\beta$ models.  Fitting these with spherical
models he found that the Hubble constant estimated from the fitting
was within $\simeq 5$\% of the true value.  We are in the process of
using N-body and smoothed particle hydrodynamics (SPH) simulations of
48 clusters to quantify the effects of complex cluster structure on
our results.

Cooling flows also affect the derived distance to the cluster,
affecting the emission weighted mean temperature and enhancing the
X-ray central surface brightness \citep[see, e.g.,][]{nagai2000}.
A characteristic cooling time for the ICM is
\begin{equation}
t_{cool} \sim \frac{3 k T_e n_{tot}}{2 \Lambda n_e \nH} 
	=  \frac{3 k T_e}{2 \Lambda n_e} \frac{\muH}{\mu_{tot}},
\label{eq:tcool1}
\end{equation}
where $\Lambda$ is the bolometric cooling function of the cluster and
all quantities are evaluated at the center of the cluster.  Cooling
flows may occur if the cooling time is less than the age of the
cluster, which we conservatively estimate to be the age of the
universe at the redshift of observation, $t_{cool} < t_{H}(z)$.  For a
flat, Einstein-de Sitter universe, the Hubble time is $t_{H}(z) =
\frac{2}{3} \Ho^{-1} (1+z)^{-3/2}$.  Both \cluster\ and \clustwo\ are
observed in the $z\approx 0.55$ universe so that $t_{H}(z=0.55) = 3
\times 10^{9}\; h^{-1} $ years.  The ratio of the cooling time to the
Hubble time for typical ICM parameters at redshift of 0.55 is then
\begin{eqnarray}
\frac{t_{cool}}{t_H(z=0.55)} \sim 19 h \left ( \frac{T_e}{8\mbox{
keV}} \right ) \left ( \frac{2.5 \times 10^{-23} \ \ecs}{\Lambda}
\right ) \nonumber\\
\times \left ( \frac{10^{-3} \mbox{ cm}^{-3}}{n_e} \right ). \ \ \ \ 
\label{eq:tcool3}
\end{eqnarray}
Using the best fit parameters we find $\Lambda= 2.8 \times 10^{-23}$
\ecs\ and $n_e=1.3 \times 10^{-2}$ cm$^{-3}$ for \cluster\ and
$\Lambda= 2.4 \times 10^{-23}$ \ecs\ and $n_e=7.0 \times 10^{-3}$
cm$^{-3}$ for \clustwo\ (the densities are determined by eliminating
\Da\ in eqs. [\ref{eq:szsignal}] and [\ref{eq:xsignal}] in favor of
\no).  This implies $t_{cool}/t_H$ ratios of $\sim 1.7$ and $\sim 2.6$
respectively.  These ratios are summarized in Table~\ref{tab:tcool}
for all three cosmologies considered in this paper.  From this simple
calculation, we do not expect cooling flows in either of these
clusters.  The X-ray radial surface brightness profiles
(Figs~\ref{fig:radprof}{\it a} and {\it b}) provide no evidence for
excess emission in the cluster core (see also \citealt{donahue1995,
neumann1997}).  As a check, we calculate $t_{cool}/t_H$ ratios for
each cluster analyzed by Mohr \etal\ (1999)\nocite{mohr1999a}.  We
check our cooling flow and non-cooling flow determinations versus
those of Peres \etal\ (1998)\nocite{peres1998} and Fabian
(1994)\nocite{fabian1994}.  Of the 45 clusters in the Mohr sample, 41
have published mass deposition rates.  We assume the cluster does not
contain a cooling flow if its mass deposition rate is consistent with
zero, otherwise it is designated as a cooling flow cluster.  We are
able to predict whether a cluster has a cooling flow or not with a
90\% success rate, suggesting that the ratio $t_{cool}/t_H$ presented
in equation~(\ref{eq:tcool3}) is a good predictor for the presence of
a cooling flow.

\begin{deluxetable}{lcccccc}
\tablewidth{0pt}
\tablecaption{Ratio of $t_{cool}/t_H(z)$ \label{tab:tcool}}
\tablehead{
\colhead{} &
\multicolumn{3}{c}{Cosmology(\Om, \Ol)}
\\
\cline{2-4}
\colhead{Cluster} &
\colhead{(1.0, 0.0)} &
\colhead{(0.3, 0.0)} &
\colhead{(0.3, 0.7)}
}
\tablecolumns{7}
\startdata
\cluster & 1.7 & 1.3 & 1.0\\
\clustwo & 2.6 & 1.9 & 1.6
\enddata
\end{deluxetable}

An isothermal analysis of a non-isothermal cluster could result in a
large distance error; moreover, an isothermal analysis of a large
cluster sample could lead to systematic errors in the derived Hubble
parameter if most clusters have similar departures from isothermality
\citep{birkinshaw1994, inagaki1995, holzapfel1997}.  The effects of
temperature variations depend on the observing technique.  For
example, PSPC X-ray constraints on the ICM distribution are very
insensitive to temperature variations for gas whose temperatures are
above 1.5 keV (see Fig.~1, \citealp{mathiesen1999}).  In principle,
SZE observations are sensitive to temperature variations, because the
SZE decrement is proportional to the projected pressure distribution
(see eq.~[\ref{eq:szsignal}]).  However, interferometric observations
of the type presented here are relatively insensitive to modest ICM
temperature variations.

Interferometric SZE observations sample the Fourier transform of the
sky brightness distribution over a limited region of the \uv\ plane.
Specifically, the radio telescope dish size imposes a minimum
separation for any two telescopes, making it impossible to sample the
Fourier transform of the cluster SZE below some minimum radius in the
\uv\ plane.  Moreover, the primary beam of the telescope defines some
effective field of view, making interferometric observations
completely insensitive to sky brightness fluctuations on any angular
scale for those regions of the sky which lie outside the field of
view.  For these reasons interferometric SZE observations are
insensitive to large angular scale variations in sky brightness.
Therefore, clusters whose core regions are approximately isothermal
and whose ICM temperatures decrease only gradually toward the virial
region pose no problems for an isothermal analysis.

We are currently analyzing mock observations of gas-dynamical cluster
simulations to explore the effects of expected temperature
distributions for our observing strategy.  These simulated clusters
exhibit X-ray merger signatures consistent with those observed in real
clusters and, presumably, they exhibit the appropriate complexities in
their temperature structure as well.  Preliminary results from this
analysis indicate that expected temperature gradients do not introduce
a large systematic error in our distance measurements.  However,
clumping within the ICM due to the common mergers of subclusters does
enhance the X-ray surface brightness by $\sim 20$\%.  This enhancement
causes X-ray gas mass estimates to be biased high by 10\%
\citep{mohr1999a}, and it results in a $\sim 20$\% underestimate of
cluster distances.  There is currently no direct observational
evidence of clumping within the ICM, but merger signatures are common
\citep{mohr1995}, and the mergers are the driving mechanism behind
these fluctuations in the simulated clusters \citep{mathiesen1999}.
We conservatively include a 20\% systematic for clumping and
departures from isothermality.

\subsubsection{Possible SZE Contaminants}
\label{subsec:sz_contam}
Undetected point sources may bias the angular diameter distance.
Point sources near the cluster center mask the SZE decrement, causing
an underestimate in the magnitude of the decrement, and therefore an
underestimate of the angular diameter distance.  While we can not rule
out point sources below our detection threshold, to estimate an upper
bound on their effects we add a point source with flux at our
detection limit near the cluster center to each data set and then fit
the new data set, not accounting for the added point source.  Such a
point source being at the cluster center is highly unlikely but
provides an upper bound to the effects of undetected point sources.
For a point source with flux density 1, 2, and 3 times the rms ($\sim
65$ $\mu$Jy) in the high-resolution ($\gsim 1,800 \lambda$) image for
\cluster, we find the magnitude of the decrement decreases by 3\%,
8\%, and 14\% respectively.  For the \clustwo\ high-resolution image
($\gsim 1,500 \lambda$), we find the decrement changes by 2\%, 10\%,
and 17\% for an on-center point source with flux density 1, 2, and 3
times the rms ($\sim 90$ $\mu$Jy) respectively.  However, we can place
more stringent constraints on contamination from undetected point
sources because we have information about the distribution of point
sources in these two fields from observations at lower frequencies.
By performing a deeper survey of these fields with our 30 GHz
receivers with an array configured for higher resolution we can lower
our point source detection threshold until the uncertainty from
undetected point sources becomes negligible.

The NVSS \citep{condon1998} detected two point sources within
400\arcsec\ from the center of \cluster.  Of these, we detect the one
that is 170\arcsec\ from the pointing center, but the point source
295\arcsec\ from the pointing center is outside the OVRO field of
view.  Sources with flux densities greater than $\sim 4\; \sigma
\approx 2$ mJy appear in their catalog.  As a more realistic upper
bound on the contamination from undetected point sources, we
extrapolate a point source with flux density equal to 4 $\sigma$ at
1.4 GHz to 28.5 GHz using the average spectral index of radio sources
in galaxy clusters $\alpha = 0.77$ \citep{cooray1998a}.  Therefore, we
place a 180 $\mu$Jy point source near the center of \cluster\ and then
fit the new image, not accounting for the additional point source.
The magnitude of the central decrement decreases by 13\% which is a
reasonable upper bound to the contamination from undetected point
sources in the \cluster\ field and similar to the constraints derived
from our own data.

\citet{moffet1989} surveyed the region around \clustwo\ at 5 GHz with
the VLA and then followed up the 5 GHz sources at 1.4 GHz and 15 GHz.
Three of their sources (10, 14, and 15) fall within the BIMA field of
view.  We detect source 15 in the BIMA data, but it falls outside the
OVRO field of view at 338\arcsec\ from the pointing center.  We
extrapolate sources 10 and 14 to 28.5 GHz from the 1.4 GHz
observations using the spectral index $\alpha = 0.77$, which is
consistent with the Moffet \& Birkinshaw result, $\alpha = 0.7$.
After correction for the primary beam, sources 10 and 14 are expected
to be 28 $\mu$Jy and 227 $\mu$Jy, respectively.  We add these two point
sources to the OVRO data placing them at their NVSS positions, perform
a model fit not accounting for them, and find a 3\% change in the
central decrement.  Moffet \& Birkinshaw searched for peaks that were
5$\sigma$ or greater.  Extrapolating the 5 GHz rms of 80 $\mu$Jy to
28.5 GHz results in a 21 $\mu$Jy rms.  Placing a 5$\sigma$ (100
$\mu$Jy) point source near the cluster center decreases the magnitude
of the central decrement by 3\%, identical to the combined effect from
sources 10 and 14.

Cluster peculiar velocities with respect to the CMB introduce an
additional CMB spectral distortion known as the kinetic SZE.  The
kinetic SZE is proportional to the thermal effect but has a different
spectral signature so it can be disentangled from the thermal SZE with
spectral SZE observations.  For a 10 keV cluster with a line-of-sight
peculiar velocity of 1000 \kms, the kinetic SZE is $\sim 11$\% of the
thermal SZE at 30 GHz.  Watkins (1997)\nocite{watkins1997} presented
observational evidence suggesting a one-dimensional rms peculiar
velocity of $\sim 300$ \kms\ for clusters, and recent simulations
found similar results \citep{colberg2000}.  With a line-of-sight
peculiar velocity of 300 \kms\ and a more typical 8 keV cluster, the
kinetic SZE is $\sim 4$\% of the thermal effect, introducing up to a
$\sim \pm 8$\% correction to the angular diameter distance computed
from one cluster.  The effects from peculiar velocities when averaged
over an ensemble of clusters should cancel, manifesting itself as an
additional statistical uncertainty similar to the effects of
asphericity.

CMB primary anisotropies have the same spectral signature as the
kinetic SZE.  Recent BIMA observations provide limits on primary
anisotropies on the scales of the observations presented here
\citep{holzapfel2000}.  We place a 95\% confidence upper limit to the
primary CMB anisotropies of $\Delta T < 22$ $\mu$K at $\ell \sim 5500$
($\sim 2\arcmin$ scales).  Thus primary CMB anisotropies are an
unimportant ($\lsim 2$\%) source of uncertainty for our observations.

\section{Discussion and Conclusions}
\label{sec:disc_concl}
We perform a maximum-likelihood joint fit to interferometric SZE and
{\it ROSAT} X-ray (PSPC and HRI) data to constrain the ICM parameters
for \cluster\ and \clustwo.  We model the ICM as a spherical,
isothermal $\beta$ model.  From this analysis we determine the
distances to be $1278^{+265}_{-298}$ Mpc and $2041^{+484}_{-514}$ Mpc
for \cluster\ and \clustwo, respectively (statistical uncertainties
only).  Together, these distances imply a Hubble constant of
\begin{equation}
\Ho = \cases{
	52 ^{+10}_{-\phn 7} \, ^{+17}_{-17} \ \ksM; &\Om=1.0, \Ol=0.0, \cr
	57 ^{+11}_{-\phn 8} \, ^{+19}_{-19} \ \ksM; &\Om=0.3, \Ol=0.0, \cr
	63 ^{+12}_{-\phn 9} \, ^{+21}_{-21} \ \ksM; &\Om=0.3, \Ol=0.7, \cr
}
\label{eq:Horesult_sys}
\end{equation}
where the uncertainties are statistical followed by systematic at 68\%
confidence.  The systematic uncertainties have been added in
quadrature and include an 8\% (4\% in \dTo) uncertainty from the
absolute calibration of the SZE data, a 10\% effective area
uncertainty for the PSPC and HRI, a 5\% uncertainty from the column
density, a 14\% ($\simeq 20/\sqrt{2}$) uncertainty due to asphericity,
a 20\% effect for our assumptions of isothermality and single-phase
gas, a 16\% (8\% in \dTo) uncertainty from undetected radio sources,
and a 6\% ($\simeq 8/\sqrt{2}$) uncertainty from the kinetic SZE.
These systematic uncertainties are summarized in
Table~\ref{tab:Hosyserr}.  The uncertainty from undetected radio
sources is the average of the maximum effects due to undetected
sources for the \cluster\ (26\%) and \clustwo\ (6\%) fields.  The
contributions from asphericity and kinetic SZE should average out for
a large sample.

Our \Ho\ determination from high-redshift clusters is consistent with
other SZE based measurements as well as the recent results from the
{\it Hubble Space Telescope} ({\it HST}) \Ho\ Key Project, which
probed the nearby universe and found $\Ho = 71 \pm 6$ \ksM\
\citep{mould2000}.  Birkinshaw (1999)\nocite{birkinshaw1999} compiled
current SZE based \Ho\ measurements and found an ensemble average of
$\sim 60$ \ksM\ independent of the chosen cosmology; where the
uncertainty is difficult to ascertain because the measurements are not
independent, many share SZE or X-ray data, and nearly all share common
absolute calibrations.

The SZE derived distances are direct, making them an interesting check
of the cosmological distance ladder.  Recent observations of masers
orbiting the nucleus of the nearby galaxy NGC4258
\citep{herrnstein1999} illustrate a method of determining direct
distances in the nearby universe.  Time delays from analysis of
gravitational lensing data from galaxy clusters are another direct
distance indicator that can probe the high redshift universe
\citep[for recent examples see][]{fassnacht1999, biggs1999,
lovell1998, barkana1997, schechter1997}.  The redshift independence of
the SZE makes it a powerful probe of clusters at high redshift.  The
combination of SZE and deep X-ray observations could provide a
valuable independent check of high redshift SN Ia results
\citep{schmidt1998, perlmutter1999}, which constrain the geometry of
the universe.

We are currently analyzing a larger sample of SZE clusters which will
reduce the statistical uncertainty as well as effects from asphericity
and the kinetic SZE.  Analysis of mock observations of simulated
clusters will also provide insight into the effects of temperature
gradients and multiphase ICMs.  With the recent launch of {\it
Chandra} and the impending launch of {\it XMM}, we will soon obtain
better $T_e$ measurements (currently a large source of observational
uncertainty; see Table~\ref{tab:Daerr}) and measure temperature
profiles.  In addition, the $\sim 2$\% absolute calibration
uncertainty of {\it Chandra} will soon replace the $\sim 10$\% {\it
ROSAT} absolute calibration uncertainty.  There is also work being
done to improve the absolute calibration at 30 GHz using the planets.
The goal is to achieve a $\lsim 1$\% absolute calibration, further
reducing the systematic uncertainties in the derived Hubble parameter.

\acknowledgments 

This work is supported by NASA LTSA grant NAG5-7986.  We thank Cheryl
Alexander, Paul Whitehouse, and Doug Huie for their help in the
construction and assembly of the receivers.  We also thank Carlo
Graziani for useful discussions.  This work would not be possible
without the help of many people at both OVRO and BIMA.  In particular
we thank S.\ Padin, S.\ Scott, D.\ Woody, J.\ Lugten, R.\ Plambeck,
and J.\ R.\ Forster.  Radio astronomy with the OVRO millimeter array
is supported by NSF grant AST 96-13717.  The BIMA millimeter array is
supported by NSF grant AST 96-13998.  EDR, LG, and SKP acknowledge
support from NASA GSRP Fellowships NGT5-50173, NGT-51201, and
NGT8-52863 respectively.  JJM is supported through Chandra Fellowship
grant PF8-1003, awarded through the Chandra Science Center.  The
Chandra Science Center is operated by the Smithsonian Astrophysical
Observatory for NASA under contract NAS8-39073.  JC acknowledges
support from a NSF-YI grant and the David and Lucile Packard
Foundation.  JPH acknowledges support from NASA LTSA grant NAG5-3432.
This research has made use of data obtained through the High Energy
Astrophysics Science Archive Research Center (HEASARC) Online Service,
provided by the NASA/Goddard Space Flight Center.  This work has also
made use of the online NVSS and FIRST catalogs operated by the NRAO as
well as the NASA/IPAC Extragalactic Database (NED) which is operated
by the Jet Propulsion Laboratory, California Institute of Technology,
under contract with the National Aeronautics and Space Administration.

\bibliographystyle{apj}
\bibliography{apj-jour,/home/piglet/reese/latex/er}

\end{document}